\documentclass[12pt,twoside, letterpaper, margin=1in]{article}
\usepackage{amssymb,amsmath,amsthm,latexsym}
\usepackage{amsfonts}
  \usepackage{amsfonts}
\usepackage{graphicx}
\usepackage{hyperref} 
\usepackage{subfigure}
\usepackage{float}
\usepackage{caption}
\usepackage{setspace}
\usepackage[margin=1in]{geometry}
\usepackage{booktabs}
\usepackage{makecell}
\usepackage{adjustbox}
\usepackage{appendix}
\usepackage[utf8]{inputenc}
\usepackage[linesnumbered,ruled,vlined]{algorithm2e}
\usepackage{tabularx} 
\usepackage{lmodern}
\usepackage{siunitx} 
\usepackage{enumitem}
\vspace{5cm}

\begin{document}
  \singlespacing
  \label{'ubf'}  
\setcounter{page}{1}                                 

\markboth {\hspace*{-9mm} \centerline{\footnotesize \sc
     }
                 }
                { \centerline                           {\footnotesize \sc  
                                                    } \hspace*{-9mm}              
               }

\vspace*{-2cm}

\begin{center}
{ 
       {\Large\textbf{A Novel SIMD-Optimized Implementation for Fast and Memory-Efficient Trigonometric Computation}}
\\
\medskip

{\sc Nikhil Dev Goyal$^{1}$, Parth Arora$^{2}$} \\

\medskip

{\footnotesize
$^{1}$ Thapar Institute of Engineering and Technology,\\
e-mail: {\it nikhil.025.goyal@gmail.com} \\

$^{2}$ Indian Institute of Technology Bombay,\\
e-mail: {\it partharora7227@gmail.com}
}
}
\end{center}

\thispagestyle{empty}

\hrulefill

\begin{abstract}  
{\footnotesize This paper proposes a novel set of trigonometric implementations which are \(5 \times\) faster than the inbuilt C++ functions. The proposed implementation is also highly memory efficient requiring no precomputations of any kind. Benchmark comparisons are done versus inbuilt functions and an optimized taylor implementation. Further, device usage estimates are also obtained, showing significant hardware usage reduction compared to inbuilt functions. This improvement could be particularly useful for low-end FPGAs or other resource-constrained devices.
}
 \end{abstract}
 \hrulefill


\section{Introduction}

Calculation of Trigonometric Functions is an essential task in fields like computer graphics, digital signal processing, scientific computing among other things. Hence the need for efficient approximation algorithms becomes crucial. A detailed survey of Trigonometric applications in engineering and technology can be found in \cite{r1}. \\
Existing methods to compute trigonometric functions include, but are not limited to: 
\begin{enumerate}[label=\roman*]
    \item Taylor series expansions (\cite{taylorser}) are general trigonometric approximations which are slow to converge and cannot be calculated on a wide definitive domain.
    \item CORDIC algorithms (\cite{cordic_1}, \cite{cordic_2}) are iterative approximation algorithms which are effiecient for hardware implementations but usually suffer from slow convergence thus requiring large number of iterations to reach the desired result.
    \item LUT (Look-Up Table) based approaches (\cite{lut_1}, \cite{lut_2}) precompute and store values at certain points of the function, often offering constant time retrievals. They suffer from the problem of high-memory usage as the size of the lookup table increases exponentially. They can also suffer from interpolation errors when combined with polynomial methods.
    \item Polynomial methods (\cite{poly1}, \cite{poly2}, \cite{poly3}, \cite{poly4}) use polynomial functions to approximate trigonometric functions. Computational efficiency of these methods depend upon the degree of the polynomial used.
\end{enumerate}
This paper proposes a novel set of trigonometric functions that are \(5 \times\) faster than its inbuilt C++ counterpart while maintaining high accuracy and requiring no precomputations, making it a suitable candidate to be used for hardware implementations like FPGA, MCU etc. This is achieved by piecewise approximation functions combined with SIMD techniques for efficient implementation in C++. Particularly, the proposed tangent function demonstrates remarkable accuracy, achieving a maximum absolute error of just 0.27 even in regions near odd multiples of $\frac{\pi}{2}$ where tangent value diverges and become very hard to estimate. This effectively addresses the limitation of Taylor series in this regard. Additionally, optimized implementations of Taylor series are obtained by SIMD techniques in C++. A detailed comparative analysis between proposed functions and taylor functions is also drawn. \\
Although SIMD optimizations are not directly available in HLS tools due to the lack of vector execution units in FPGA, optimization techniques like loop unrolling and pipelining are realized.\\
Device usage estimates are drawn using Vitis High-Level Synthesis (HLS) 2024.2 targeting the \textbf{Xilinx Artix-7 XC7A35T-CPG236-1 FPGA}, demonstrating significant reductions in hardware resources such as DSPs, LUTs, and flip-flops (FFs) as compared to inbuilt libraries. Artix-7 XC7A35T-CPG236-1 was chosen for benchmarking to effectively evaluate the memory efficiency of the proposed formulas, particularly in resource-limited environments.

\section{Proposed formulas}
\label{sec:proposed}
The proposed formulas are obtained by performing linear interpolation on $sin\,x$ and $cos\,x$ for uniform ranges of $\frac{\pi}{12}$ and subsequently dividing them for corresponding ranges to get expressions for $tan\,x$. A significant increase in accuracy is observed in the resulting $tan\,x$ functions. Subsequently, these functions are used to arrive at the final expressions for $sin\,x$ and $cos\,x$ using standard trigonometric identities. The range of $\frac{\pi}{12}$ was chosen to balance the number of piecewise functions while maintaining high accuracy.\\
The proposed formulas are listed below; refer to \autoref{appendix:proofs} for detailed mathematical derivations for them.

\subsection*{Sine Function}
\[
\sin x =
\begin{cases}
\displaystyle \frac{630.25 x}{\sqrt{400502.37 x^2 +-72421.86 x + 399424}}, & 0 \le x < \dfrac{\pi}{12}, \\[3ex]
\displaystyle \frac{572.95 x + 10.0}{\sqrt{380805.53 x^2 + -289687.46 x +431749.0}}, & \dfrac{\pi}{12} \le x < \dfrac{\pi}{6}, \\[3ex]
\displaystyle \frac{343.77 x +  46.0}{\sqrt{200251.18 x^2 + -278342.89 x + 294797.0}}, & \dfrac{\pi}{6} \le x < \dfrac{\pi}{4}, \\[3ex]
\displaystyle \frac{572.95 x + 217.0}{\sqrt{883074.90 x^2 + -1616428.53 x + 1614593.0}}, & \dfrac{\pi}{4} \le x < \dfrac{\pi}{3}, \\[3ex]
\displaystyle \frac{229.18 x + 297.0}{\sqrt{380805.53 x^2 + -906648.41 x + 916309.0}}, & \dfrac{\pi}{3} \le x < \dfrac{5\pi}{12}, \\[3ex]
\displaystyle \frac{57.29 x + 542.0}{\sqrt{400502.37 x^2 + -1185793.45 x + 1273864.0}}, & \dfrac{5\pi}{12} \le x \le \dfrac{\pi}{2}.
\end{cases}
\]

\subsection*{Cosine Function}
\[
\cos x =
\begin{cases}
\displaystyle \frac{-57.29 x + 632.0}{\sqrt{400502.37 x^2 + -72421.86 x + 399424.0}}, & 0 \le x < \dfrac{\pi}{12}, \\[3ex]
\displaystyle \frac{-229.18 x + 657.0}{\sqrt{380805.53 x^2 + -289687.46 x + 431749.0}}, & \dfrac{\pi}{12} \le x < \dfrac{\pi}{6}, \\[3ex]
\displaystyle \frac{-286.47 x + 541.0}{\sqrt{200251.18 x^2 + -278342.89 x + 294797.0}}, & \dfrac{\pi}{6} \le x < \dfrac{\pi}{4}, \\[3ex]
\displaystyle \frac{-744.84 x + 1252.0}{\sqrt{883074.90 x^2 + -1616428.53 x + 1614593.0}}, & \dfrac{\pi}{4} \le x < \dfrac{\pi}{3}, \\[3ex]
\displaystyle \frac{-572.95 x + 910.0}{\sqrt{380805.53 x^2 + -906648.41 x + 916309.0}}, & \dfrac{\pi}{3} \le x < \dfrac{5\pi}{12}, \\[3ex]
\displaystyle \frac{-630.25 x + 990.0}{\sqrt{400502.37 x^2 + -1185793.45 x + 1273864.0}}, & \dfrac{5\pi}{12} \le x \le \dfrac{\pi}{2}.
\end{cases}
\]

\subsection*{Tangent Function}
\[
\tan x =
\begin{cases}
\displaystyle \frac{630.25 x}{-57.29 x + 632.0}, & 0 \le x < \dfrac{\pi}{12}, \\[3ex]
\displaystyle \frac{572.95 x + 10.0}{-229.18 x + 657.0}, & \dfrac{\pi}{12} \le x < \dfrac{\pi}{6}, \\[3ex]
\displaystyle \frac{343.77 x + 46.0}{-286.47 x + 541.0}, & \dfrac{\pi}{6} \le x < \dfrac{\pi}{4}, \\[3ex]
\displaystyle \frac{572.95 x + 217.0}{-744.84 x + 1252.0}, & \dfrac{\pi}{4} \le x < \dfrac{\pi}{3}, \\[3ex]
\displaystyle \frac{229.18 x + 297.0}{-572.95 x + 910.0}, & \dfrac{\pi}{3} \le x < \dfrac{5\pi}{12}, \\[3ex]
\displaystyle \frac{57.29 x + 542.0}{-630.25 x + 990.0}, & \dfrac{5\pi}{12} \le x < \dfrac{\pi}{2}.
\end{cases}
\]
\vspace{1mm}
{\small \textbf{Note:} Coefficients are truncated to two decimal places for presentation in this paper, full precision was used for the C++ implementation\cite{gitcode}.}

\section{Implementation}
The proposed formulas are implemented in C++ and extensively optimized using SIMD techniques. Additionally, fast inverse square root (FISR)\cite{fisr} algorithm is used to avoid the use of the inbuilt square root function which is computationally quite expensive. FISR algorithm is also used to calculate reciprocal in the implementation of the proposed tan function to eliminate the need for division operations entirely.\\
Since the proposed formulas are defined for $[0, \frac{\pi}{2}]$, angles will be first efficiently reduced to this range along with the resulting sign. Concise Algorithms for the functions used are provided below, and detailed implementation can be found at \cite{gitcode}.\\

\begin{algorithm}[H]
\caption{fastInverseSqrt\_SIMD(number)}
\KwIn{number --- a \texttt{\_\_m256d} vector of 4 doubles}
\KwOut{A \texttt{\_\_m256d} vector approximating the inverse square-root of \texttt{number}}
Set \texttt{threeHalfs} $\gets$ broadcast(1.5)\;
Compute \texttt{x2} $\gets$ number $\times$ 0.5\;
\textbf{// Reinterpret the double vector as 64-bit integers}\;
Let \texttt{i} $\gets$ reinterpret \texttt{number} as \texttt{\_\_m256i}\;
Update \texttt{i} $\gets$ (magic constant \texttt{0x5fe6eb50c7b537a9}) $-$ (right-shift \texttt{i} by 1)\;
Reinterpret \texttt{i} as a double vector and store in \texttt{y}\;
\textbf{// Newton–Raphson iteration}\;
Compute \texttt{y} $\gets$ y $\times$ (threeHalfs $-$ (x2 $\times$ y $\times$ y))\;
\Return y\;
\end{algorithm}

\begin{algorithm}[H]
\caption{sin\_helper\_SIMD(a, b, x, y, z, ang)}
\KwIn{Coefficients \texttt{a}, \texttt{b}, \texttt{x}, \texttt{y}, \texttt{z} (each as \texttt{\_\_m256d}) and angle vector \texttt{ang} (\texttt{\_\_m256d})}
\KwOut{A \texttt{\_\_m256d} vector approximating the sine function}
Compute \texttt{poly} $\gets$ a $\times$ ang $+$ b\;
Compute \texttt{ang2} $\gets$ ang $\times$ ang\;
Compute \texttt{inner} $\gets$ x $\times$ ang2 $+$ y $\times$ ang $+$ z\;
Compute \texttt{invSqrt} $\gets$ fastInverseSqrt\_SIMD(inner)\;
\Return poly $\times$ invSqrt\;
\end{algorithm}

\begin{algorithm}[H]
\caption{cos\_helper\_SIMD(c, d, x, y, z, ang)}
\KwIn{Coefficients \texttt{c}, \texttt{d}, \texttt{x}, \texttt{y}, \texttt{z} (each as \texttt{\_\_m256d}) and angle vector \texttt{ang} (\texttt{\_\_m256d})}
\KwOut{A \texttt{\_\_m256d} vector approximating the cosine function}
Compute \texttt{poly} $\gets$ c $\times$ ang $+$ d\;
Compute \texttt{ang2} $\gets$ ang $\times$ ang\;
Compute \texttt{inner} $\gets$ x $\times$ ang2 $+$ y $\times$ ang $+$ z\;
Compute \texttt{invSqrt} $\gets$ fastInverseSqrt\_SIMD(inner)\;
\Return poly $\times$ invSqrt\;
\end{algorithm}

\begin{algorithm}[H]
\caption{tan\_helper\_SIMD(a, b, c, d, ang)}
\KwIn{Coefficients \texttt{a}, \texttt{b}, \texttt{c}, \texttt{d} (each as \texttt{\_\_m256d}) and angle vector \texttt{ang} (\texttt{\_\_m256d})}
\KwOut{A \texttt{\_\_m256d} vector approximating the tangent function}
Compute \texttt{poly} $\gets$ a $\times$ ang $+$ b\;
Compute \texttt{temp} $\gets$ fastInverseSqrt\_SIMD(c $\times$ ang $+$ d)\;
\Return poly $\times$ (temp $\times$ temp)\;
\end{algorithm}
The proposed sin and proposed cos use the same SIMD based approach differing only in polynomial parameters thus only the proposed sin algorithm is discussed below.\\

\begin{algorithm}[H]
\caption{proposed\_sin(ang)}
\KwIn{Scalar angle \texttt{ang} (double)}
\KwOut{Approximation of $\sin(\texttt{ang})$ (double)}
\textbf{// Broadcast the angle into all SIMD lanes}\;
Set \texttt{ang\_vec} $\gets$ \_mm256\_set1\_pd(ang)\;
\textbf{// Reduce the angle using SIMD branchless reduction for sine}\;
Compute \texttt{reduced} $\gets$ reduce\_angle\_sin\_SIMD(ang\_vec)\;
\textbf{// Extract representative reduced angle for parameter selection}\;
Store the first element of \texttt{reduced.red} into scalar variable \texttt{red\_scalar}\;
\textbf{// Since the proposed functions occur at regular intervals a precomputed array for coefficients is stored}\;
\textbf{// Determine the parameter index based on the reduced angle}\;
Compute \texttt{idx} $\gets$ floor(\texttt{red\_scalar} / 0.261)\;
Retrieve parameters \texttt{p} from array \texttt{sinparams} at index \texttt{idx}\;
\textbf{// Broadcast polynomial coefficients into SIMD vectors}\;
Set \texttt{a} $\gets$ \_mm256\_set1\_pd(p.a), \quad \texttt{b} $\gets$ \_mm256\_set1\_pd(p.b)\;
Set \texttt{x} $\gets$ \_mm256\_set1\_pd(p.x), \quad \texttt{y} $\gets$ \_mm256\_set1\_pd(p.y), \quad \texttt{z} $\gets$ \_mm256\_set1\_pd(p.z)\;
\textbf{// Compute the sine approximation using the SIMD helper}\;
Compute \texttt{res\_vec} $\gets$ sin\_helper\_SIMD(a, b, x, y, z, reduced.red)\;
\textbf{// Apply the sign correction from the angle reduction}\;
Multiply \texttt{res\_vec} by \texttt{reduced.sign} to obtain the corrected result\;
\textbf{// Extract and return the final scalar result}\;
Store the first element of \texttt{res\_vec} into scalar variable \texttt{result}\;
\Return \texttt{result}\;
\end{algorithm}

\begin{algorithm}[H]
\caption{proposed\_tan(ang)}
\KwIn{Scalar angle \texttt{ang} (double)}
\KwOut{Approximation of $\tan(\texttt{ang})$ (double)}
\textbf{// Broadcast the angle into all SIMD lanes}\;
Set \texttt{ang\_vec} $\gets$ \_mm256\_set1\_pd(ang)\;
\textbf{// Reduce the angle using SIMD branchless reduction}\;
Compute \texttt{reduced} $\gets$ reduce\_angle\_tan\_SIMD(ang\_vec)\;
\textbf{// Extract representative reduced angle for parameter selection}\;
Store first element of \texttt{reduced.red} into \texttt{red\_scalar}\;
\textbf{// Check for near $\pi/2$ to avoid division-by-zero}\;
\If{$|red\_scalar - (\pi/2)| < 1\times10^{-3}$}{
    \Return $\infty$\;
}
Compute \texttt{idx} $\gets$ floor(\texttt{red\_scalar} / 0.261)\;
Retrieve \texttt{p} from \texttt{tanparams[idx]}\;
\textbf{// Broadcast parameters into SIMD vectors}\;
Set \texttt{a} $\gets$ \_mm256\_set1\_pd(p.a), \quad \texttt{b} $\gets$ \_mm256\_set1\_pd(p.b)\;
Set \texttt{c} $\gets$ \_mm256\_set1\_pd(p.c), \quad \texttt{d} $\gets$ \_mm256\_set1\_pd(p.d)\;
\textbf{// Compute tangent using the SIMD helper}\;
Compute \texttt{res\_vec} $\gets$ tan\_helper\_SIMD(a, b, c, d, reduced.red)\;
Apply sign correction: \texttt{res\_vec} $\gets$ res\_vec $\times$ reduced.sign\;
\textbf{// Extract and return the result}\;
Store first element of \texttt{res\_vec} into \texttt{result}\;
\Return \texttt{result}\;
\end{algorithm}

The standard algorithm of Taylor series was further optimized by the usage of SIMD techniques leading to significant improvement in its speed. For reference, additional approaches to Taylor series optimizations can be found in (\cite{tayloropt1} \cite{tayloropt2}). \\
The C++ implementation was compiled using the following command:
\begin{verbatim}
g++ -std=c++17 -mavx2 -mfma -O3 -march=native file.cpp -o output.exe
\end{verbatim}
{\small \textbf{Note: The complete C++ code for the proposed functions and the optimized Taylor series implementation is available at \cite{gitcode}}}.

\section{Benchmarking and Results}

For benchmarking 1000 random angles were generated between $-10^6$ and $10^6$ using a fixed seed (Seed  = 2024873) to ensure reproducibility of results. Each angle was computed $10^5$ times to ensure the robustness and reliability of the results. Any mentions of 'Average' in further given statistics refer to normalization over the 1000 angles.\\
For clarity the benchmarking flow is explained in the algorithm below.

\begin{algorithm}[H]
\caption{run\_benchmark(func\_name, std\_func, my\_func, num\_angles=1000, iterations=1e5)}
\KwIn{
  \texttt{func\_name}: Name of the function\;
  \texttt{std\_func}, \texttt{my\_func}: Function pointers\;
  \texttt{num\_angles}: Number of test angles (default 1000)\;
  \texttt{iterations}: Number of iterations per test (default 1e5)
}
\KwOut{BenchmarkResponse containing error metrics, execution times, and win count}
\vspace{1mm}
\textbf{// Initialization}\;
Set \texttt{results} as an empty list\;
Set \(\epsilon \gets 10^{-12}\)\;
Set \texttt{wins} \(\gets 0\)\;
\vspace{1mm}
\textbf{// Run Benchmark}\;
\For{\(i \gets 0\) \KwTo num\_angles - 1}{
  \(angle \gets\) random value in \([-10^6,10^6]\)\;
  \(std\_val \gets\) std\_func(\(angle\))\;
  \(my\_val \gets\) my\_func(\(angle\))\;
  \(abs\_err \gets |\,my\_val - std\_val\,|\)\;
  \(rel\_err \gets \begin{cases} 
      abs\_err/|std\_val|, & \text{if } |std\_val|>\epsilon \\
      0, & \text{otherwise}
    \end{cases}\)\;
  \(std\_time \gets\) measure\_execution\_time(std\_func, \(angle\), iterations)\;
  \(my\_time \gets\) measure\_execution\_time(my\_func, \(angle\), iterations)\;
  \If{\(my\_time < std\_time\)}{
    Increment \texttt{wins}\;
  }
  Append \((angle, std\_val, my\_val, abs\_err, rel\_err, std\_time, my\_time)\) to \texttt{results}\;
}
\vspace{1mm}
\textbf{// Compute Overall Statistics}\;
Compute averages, maximums, and standard deviations from \texttt{results}\;
Record total execution times and the win count \texttt{wins}\;
\vspace{1mm}
\Return BenchmarkResponse containing overall statistics and win count\;
\end{algorithm}

\subsection{Benchmarking Results}

Before presenting the results, the explanations of some less-obvious performance metrics are provided \\
\[
\text{Mean Speed-up Ratio} = \frac{\sum_{i=1}^{1000} \frac{\text{std\_time}(angle_i)}{\text{obs\_func\_time}(angle_i)}}{1000}
\]

where:  
\begin{itemize}
    \item $\text{std\_time}(angle_i)$ is the execution time of the standard (inbuilt) function for angle $angle_i$.
    \item $\text{obs\_func\_time}(angle_i)$ is the execution time of the observed (proposed/taylor) function for angle $angle_i$.
\end{itemize}

\subsubsection{Sin}

\begin{table}[H]
  \centering
  \caption{Sin Benchmark Results vs Inbuilt (\textit{double datatype})}
  \small
  \begin{tabular}{lccccc}
    \toprule
    Metric & Proposed & Taylor-5 & Taylor-5 & Taylor-7 & Taylor-9 \\ 
           &    sin     & (without SIMD) & (with SIMD) & (with SIMD) & (with SIMD) \\
    \midrule
    \textbf{Mean Speed-up Ratio} & \textbf{5.49} & \textbf{0.30} & \textbf{5.83} & \textbf{5.69} & \textbf{5.75} \\
    Wins*                   & 1000   & 0          & 1000       & 1000       & 1000\\
    Abs Max Error          & 0.001532   & 0.000003   & 0.000003   & 0.000000   & 0.000000 \\
    Avg Abs Error          & 0.000613   & 0.000000   & 0.000000   & 0.000000   & 0.000000 \\
    Avg Abs Std Dev        & 0.000520   & 0.000001   & 0.000001   & 0.000000   & 0.000000 \\
    Rel Max Error          & 0.003866   & 0.000003   & 0.000003   & 0.000000   & 0.000000 \\
    Avg Rel Error          & 0.001000   & 0.000000   & 0.000000   & 0.000000   & 0.000000 \\
    Avg Rel Std Dev        & 0.000627   & 0.000001   & 0.000001   & 0.000000   & 0.000000 \\
    \bottomrule
  \end{tabular}
  \vspace{0mm} \\
  \hspace{-1.3cm} {\footnotesize \textit{*Wins is the number of times a solution performed better than inbuilt out of 1000 angles}}
  \label{tab:sine-transposed}
\end{table}

\begin{figure}[H]
    \centering
    \subfigure[Speed-up Ratio vs Freq. for Proposed Sin]{
        \includegraphics[width=0.48\textwidth]{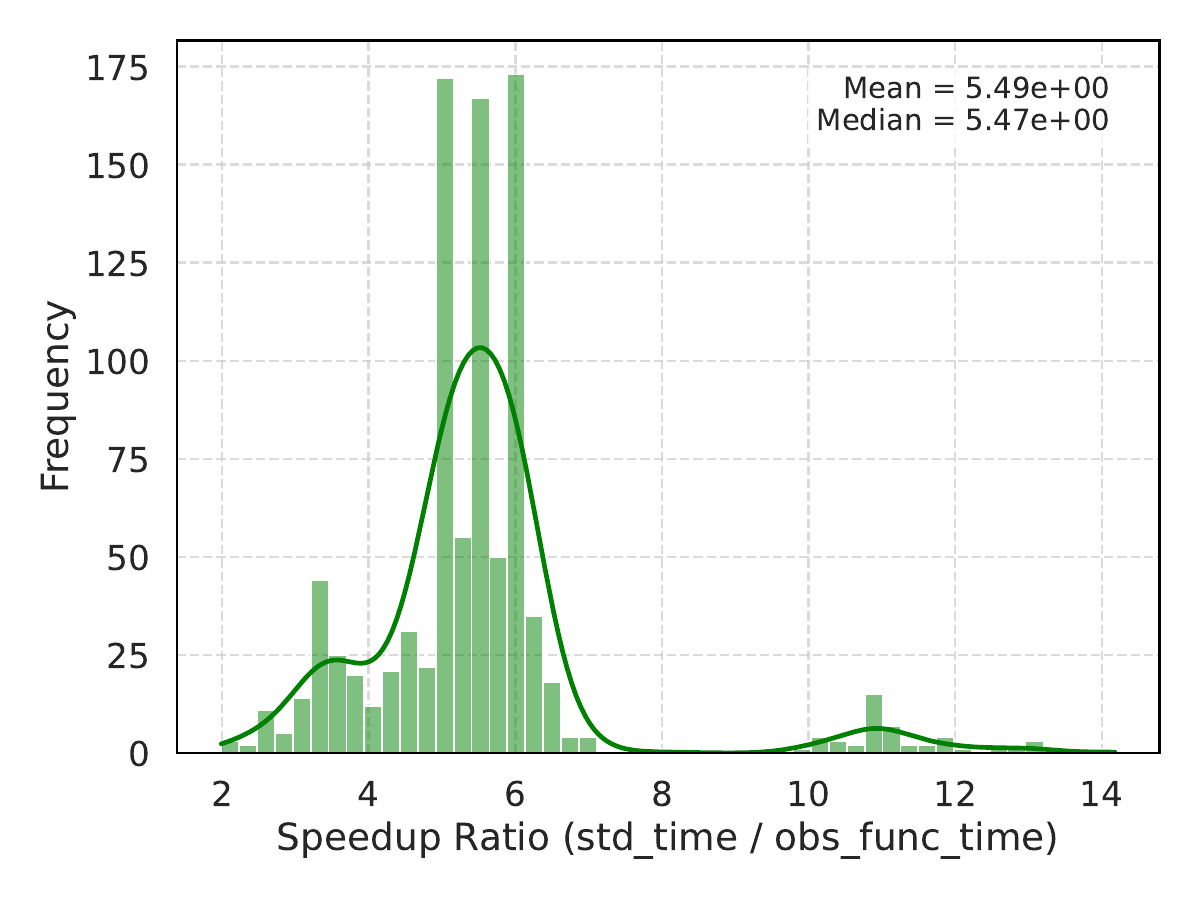}
    }\hfill
    \subfigure[Comparison of Speed-up Ratio for SIMD and Non-SIMD versions of Taylor Sin]{
        \includegraphics[width=0.48\textwidth]{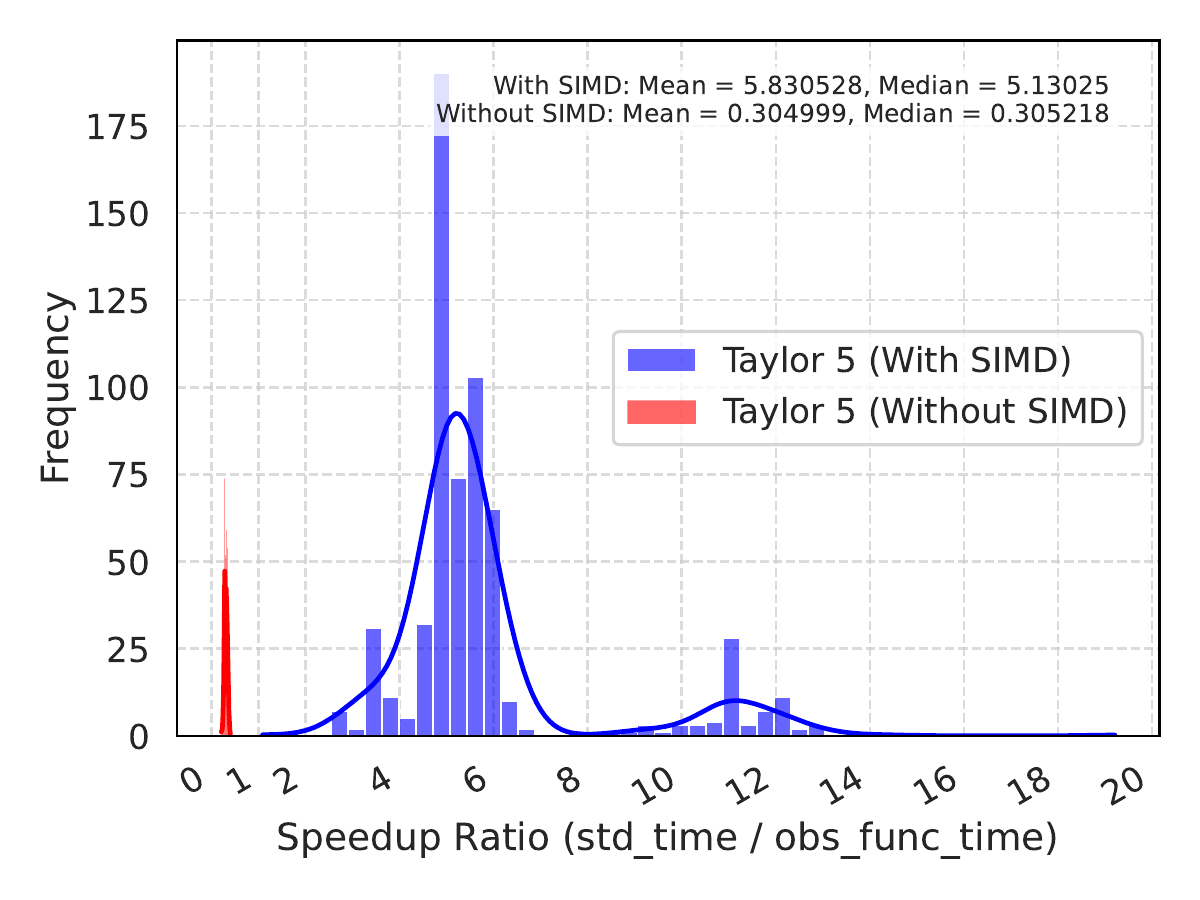}
    }
    \caption{Angle-wise Distribution of Speed-up Ratios for sin}
    \label{speed_up_ratio_sin}
\end{figure}

\begin{figure}[H]
    \centering
    \subfigure[Rel Error vs Freq. for Proposed Sin]{
        \includegraphics[width=0.48\textwidth]{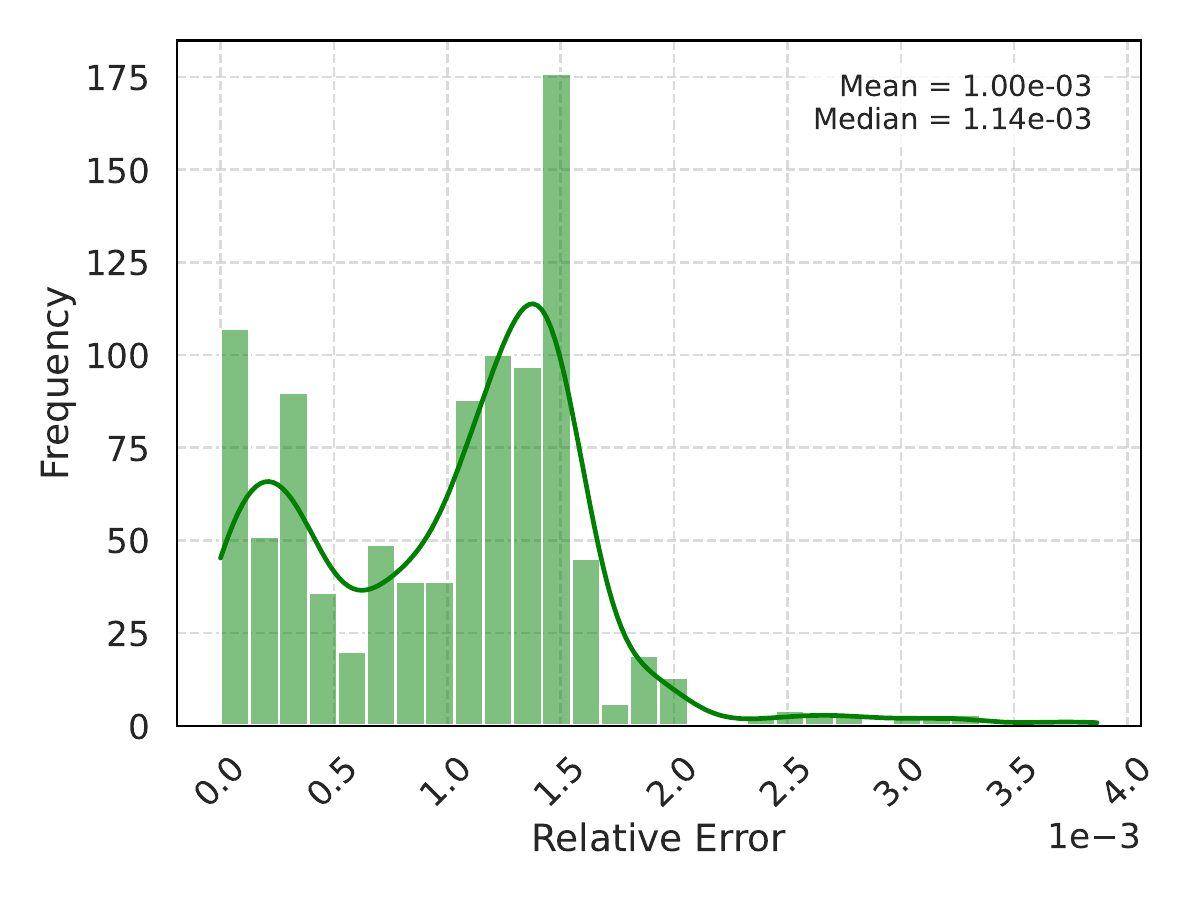}
        \label{rel_error_mysin}
    }\hfill
    \subfigure[Rel Error vs Freq. for 5th Order Taylor Sin]{
        \includegraphics[width=0.48\textwidth]{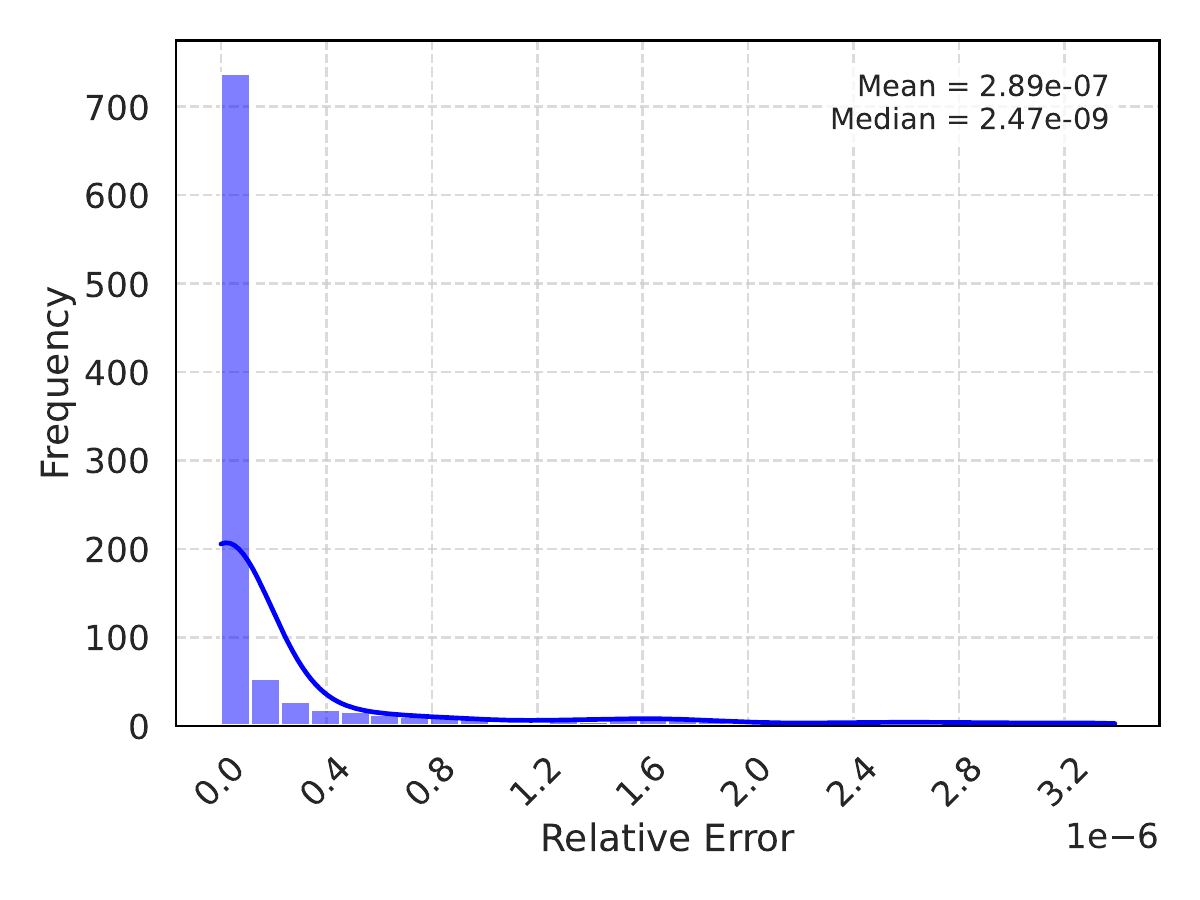}
        \label{rel_error_taylor_sin}
    }
    \caption{Angle-wise Distribution of Relative error for sin}
    \label{rel_error_sin}
\end{figure}

It can be seen from Table \ref{tab:sine-transposed} that the proposed sin function has a mean speed-up ratio of 5.49 which means that on average it was 5.49 times faster than the inbuilt library function. The Taylor series functions gain a significant improvement in computational speed with the application of SIMD techniques. Notably, the 5th-order Taylor function achieves an impressive mean speed-up ratio of 5.83. Furthermore, the distribution of the speed-up ratio across angles can be seen in Figure \ref{speed_up_ratio_sin}. 
The proposed function maintains a high accuracy with the maximum relative error bounded by 0.003866. Taylor series shows exceptional accuracy with near-zero error statistics across all metrics. Furthermore, the distribution of relative error across angles can be seen in Figure \ref{rel_error_sin}.\\
It can be concluded that both the proposed and taylor functions are equally efficient, while both being much better than the inbuilt counterpart.

{\small\textbf{Note : Angles where observed functions took 0 ns for computation have been omitted from speed-up ratio distribution graphs.}}

\subsubsection{Cos}

\begin{table}[H]
  \centering
  \caption{Cos Benchmark Results vs Inbuilt(\textit{double datatype})}
  \small
  \begin{tabular}{lccccc}
    \toprule
    Metric & Proposed & Taylor-5 & Taylor-5 & Taylor-7 & Taylor-9 \\ 
           &    cos    & (without SIMD) & (with SIMD) & (with SIMD) & (with SIMD) \\
    \midrule
    \textbf{Mean Speed-up Ratio} & \textbf{5.52} & \textbf{0.69} & \textbf{5.82} & \textbf{5.80} & \textbf{5.68} \\
    Wins*                   & 1000   & 1         & 1000       & 1000       & 1000\\
    Abs Max Error          & 0.001532   & 0.000024   & 0.000024   & 0.000000   & 0.000000 \\
    Avg Abs Error          & 0.000617   & 0.000002   & 0.000002   & 0.000000   & 0.000000 \\
    Avg Abs Std Dev        & 0.000527   & 0.000005   & 0.000005   & 0.000000   & 0.000000 \\
    Rel Max Error          & 0.003667   & 0.003805   & 0.003805   & 0.000001   & 0.000000 \\
    Avg Rel Error          & 0.000970   & 0.000055   & 0.000055   & 0.000000   & 0.000000 \\
    Avg Rel Std Dev        & 0.000637   & 0.000303   & 0.000303   & 0.000000   & 0.000000 \\
    \bottomrule
  \end{tabular}
  \vspace{0mm} \\
  \hspace{-1.3cm} {\footnotesize \textit{*Wins is the number of times a solution performed better than inbuilt out of 1000 angles}}
  \label{tab:cosine-transposed}
\end{table}

A similar trend is observed for cos as seen with sin, which is clearly demonstrated through the accompanying graphs and tables.\\

\begin{figure}[H]
    \centering
    \subfigure[Speed-up Ratio vs Freq. for Proposed Cos]{
        \includegraphics[width=0.48\textwidth]{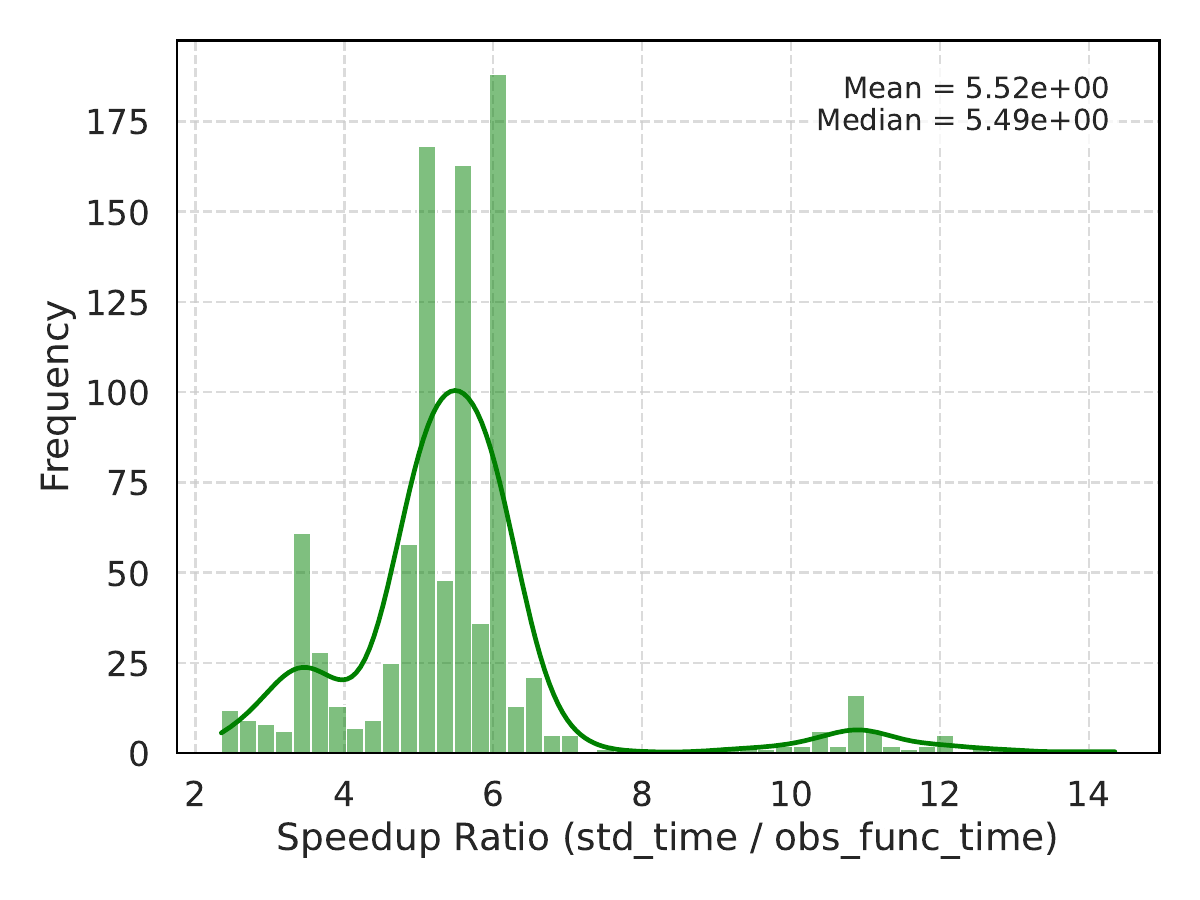}
    }\hfill
    \subfigure[Comparison of Speed-up Ratio for SIMD and Non-SIMD versions of Taylor Cos]{
        \includegraphics[width=0.48\textwidth]{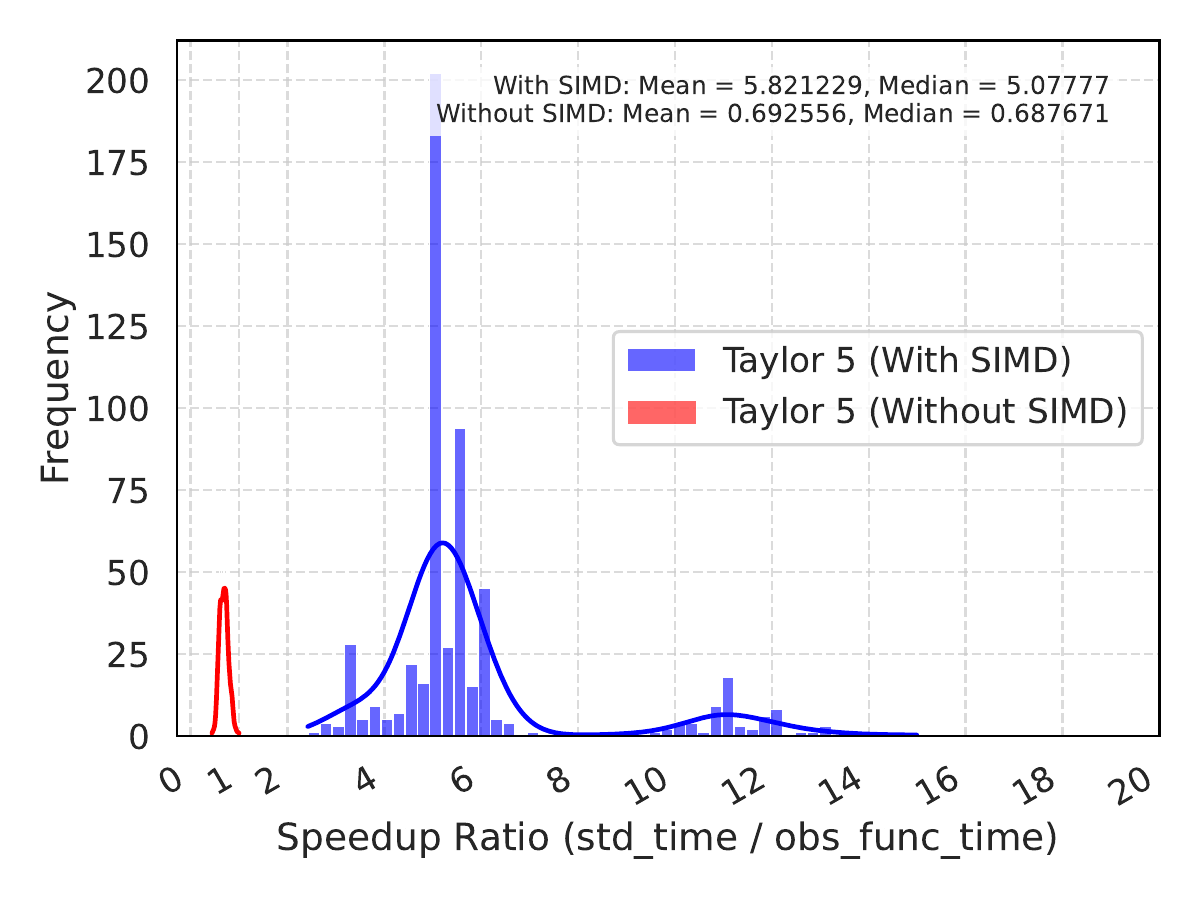}
    }
    \caption{Angle-wise Distribution of Speed-up Ratios for cos}
\end{figure}

\begin{figure}[H]
    \centering
    \subfigure[Rel Error vs Freq. for Proposed Cos]{
        \includegraphics[width=0.48\textwidth]{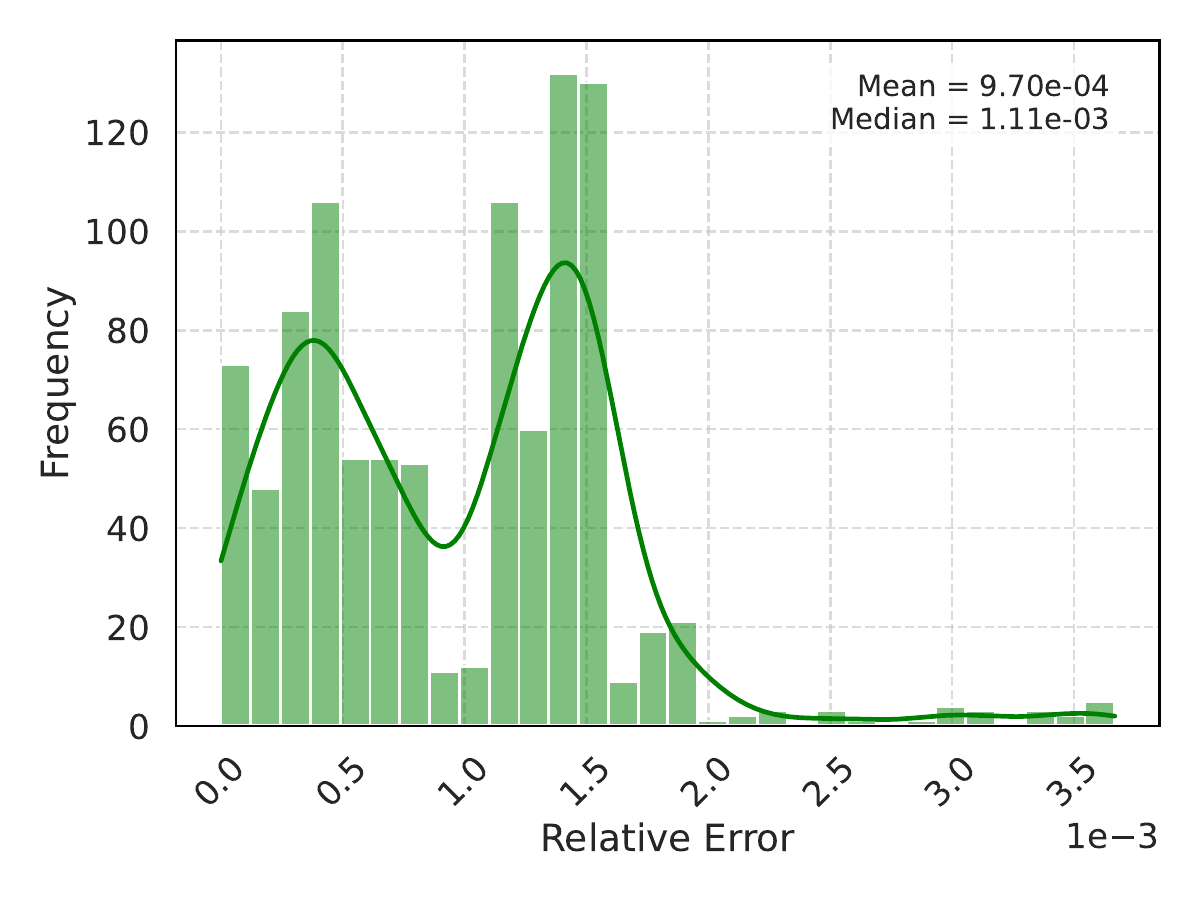}
    }\hfill
    \subfigure[Rel Error vs Freq. for 5th Order Taylor Cos]{
        \includegraphics[width=0.48\textwidth]{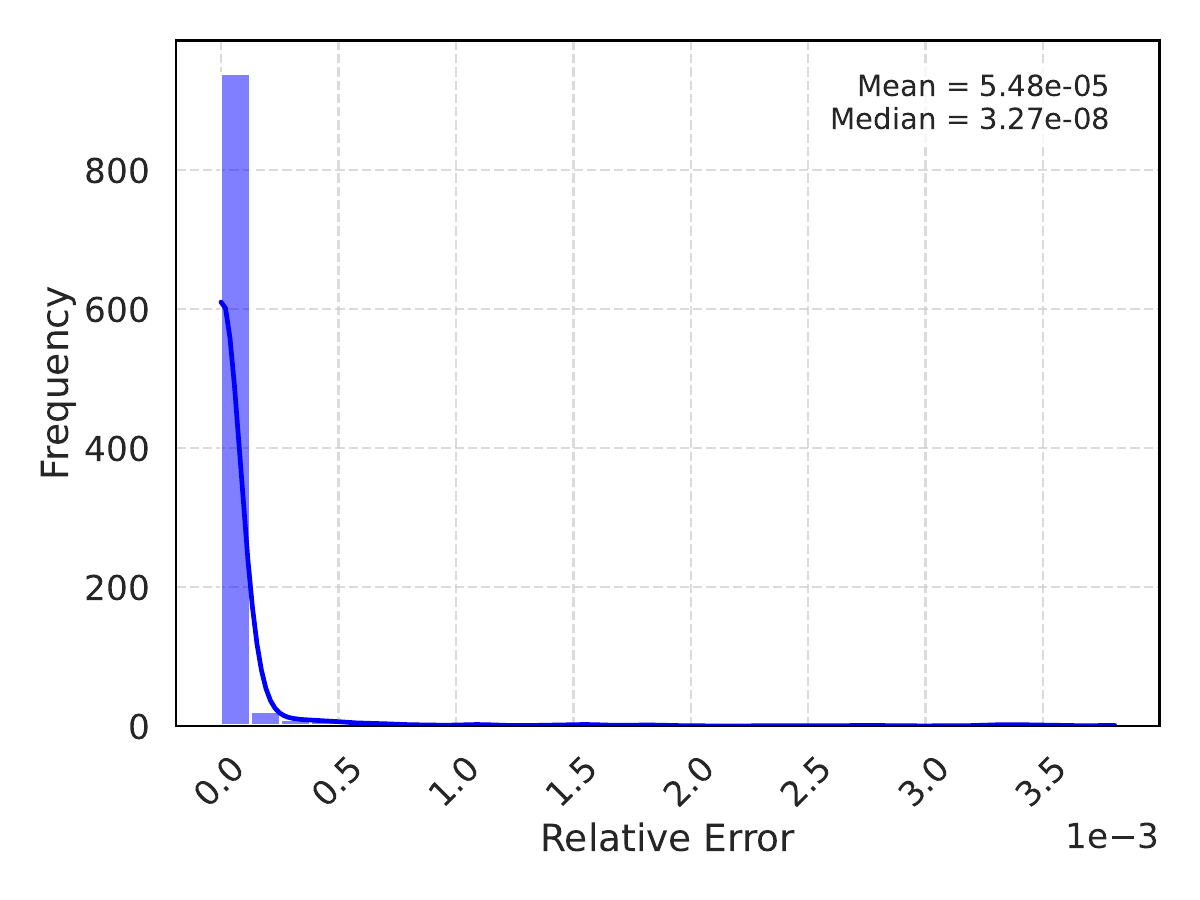}
    }
    \caption{Angle-wise Distribution of Relative error for cos}
\end{figure}

\subsubsection{Tan}

\begin{table}[H]
  \centering
  \caption{Tan Benchmark Results vs Inbuilt(\textit{double datatype})}
  \small
  \begin{tabular}{lccccc}
    \toprule
    Metric & Proposed & Taylor-5 & Taylor-5. & Taylor-7 & Taylor-9 \\ 
           &    tan     & (without SIMD) & (with SIMD) & (with SIMD) & (with SIMD) \\
    \midrule
    \textbf{Mean Speed-up Ratio} & \textbf{4.44} & \textbf{0.24} & \textbf{4.52} & \textbf{4.69} & \textbf{4.33} \\
    Wins*                   & 1000   & 0         & 1000       & 1000       & 1000\\
    Abs Max Error          & 0.273094   & 153.495978   & 153.496539   & 151.068956   & 148.681759 \\
    Avg Abs Error          & 0.008498   & 2.182710   & 2.182765   & 1.992607   & 1.846636   \\
    Avg Abs Std Dev        & 0.022166   & 12.183232  & 12.183308  & 11.860837  & 11.557965  \\
    Rel Max Error          & 0.006381   & 0.959024   & 0.959027   & 0.943860   & 0.928945   \\
    Avg Rel Error          & 0.002236   & 0.084375   & 0.084383   & 0.062845   & 0.050181   \\
    Avg Rel Std Dev        & 0.001454   & 0.193764   & 0.193771   & 0.169863   & 0.152744   \\
    \bottomrule
  \end{tabular}
  \vspace{0mm} \\
  \hspace{-1.3cm} {\footnotesize \textit{*Wins is the number of times a solution performed better than inbuilt out of 1000 angles}}
  \label{tab:tangent-transposed}
\end{table}

The proposed tan function has an impressive mean speed-up ratio of 4.44. Similar to previous trends, taylor series has been seen to receive a massive improvement in speed compared to its non-SIMD counterpart. All orders of taylor series are observed to have impressive speed performance with no major drop off across increasing orders. However, taylor series can been seen to have instability across angles in the vicinity of odd multiples of $\frac{\pi}{2}$, leading to significant absolute as well as relative errors across all orders. As discussed in the Introduction, the proposed tan formula handles this efficiently as it maintains a high accuracy across all input domains. As can be seen from Table \ref{tab:tangent-transposed}, for the same 1000 randomly generated angles the maximum absolute error reached by proposed tan is 0.27 while the maximum absolute error reaches 148.68 even considering the 9th-order taylor function. Considering the inherent difficulty of approximating tangent functions near singularities, the stability demonstrated by the proposed tan function is impressive.

\begin{figure}[H]
    \centering
    \subfigure[Speed-up Ratio vs Freq. for Proposed Tan]{
        \includegraphics[width=0.48\textwidth]{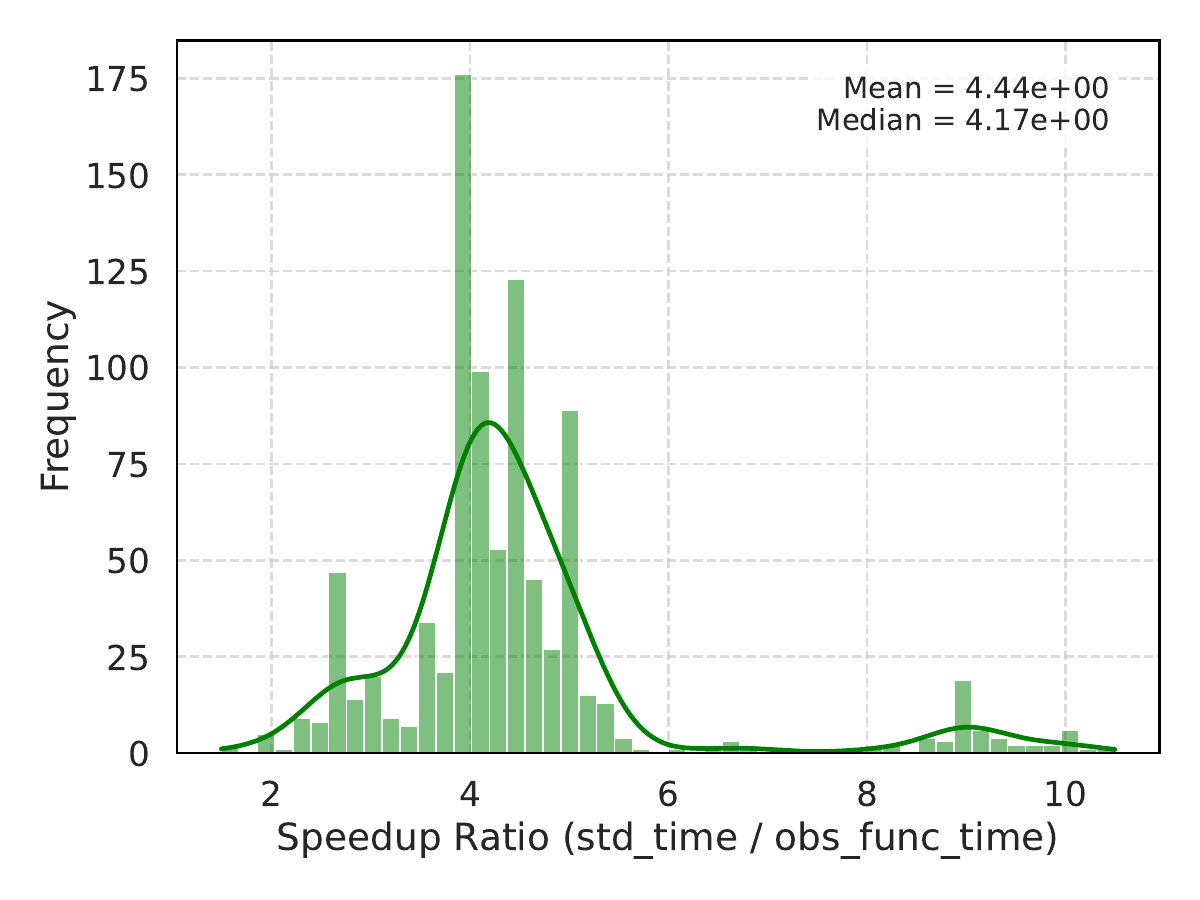}
    }\hfill
    \subfigure[Comparison of Speed-up Ratio for SIMD and Non-SIMD versions of Taylor Tan]{
        \includegraphics[width=0.48\textwidth]{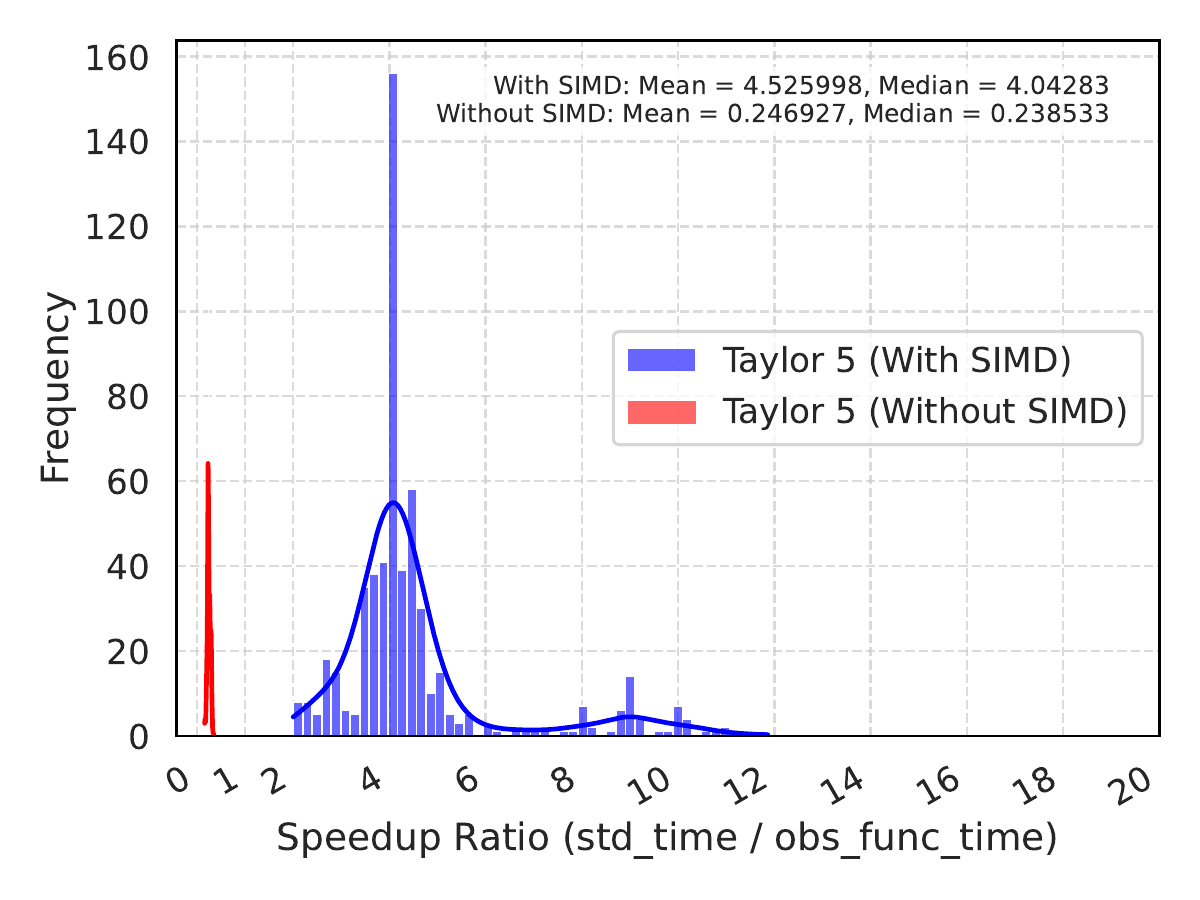}
    }
    \caption{Angle-wise Distribution of Speed-up Ratios for tan}
    \label{speed_up_ratio_tan}
\end{figure}

\begin{figure}[H]
    \centering
    \subfigure[Rel Error vs Freq. for Proposed Tan]{
        \includegraphics[width=0.48\textwidth]{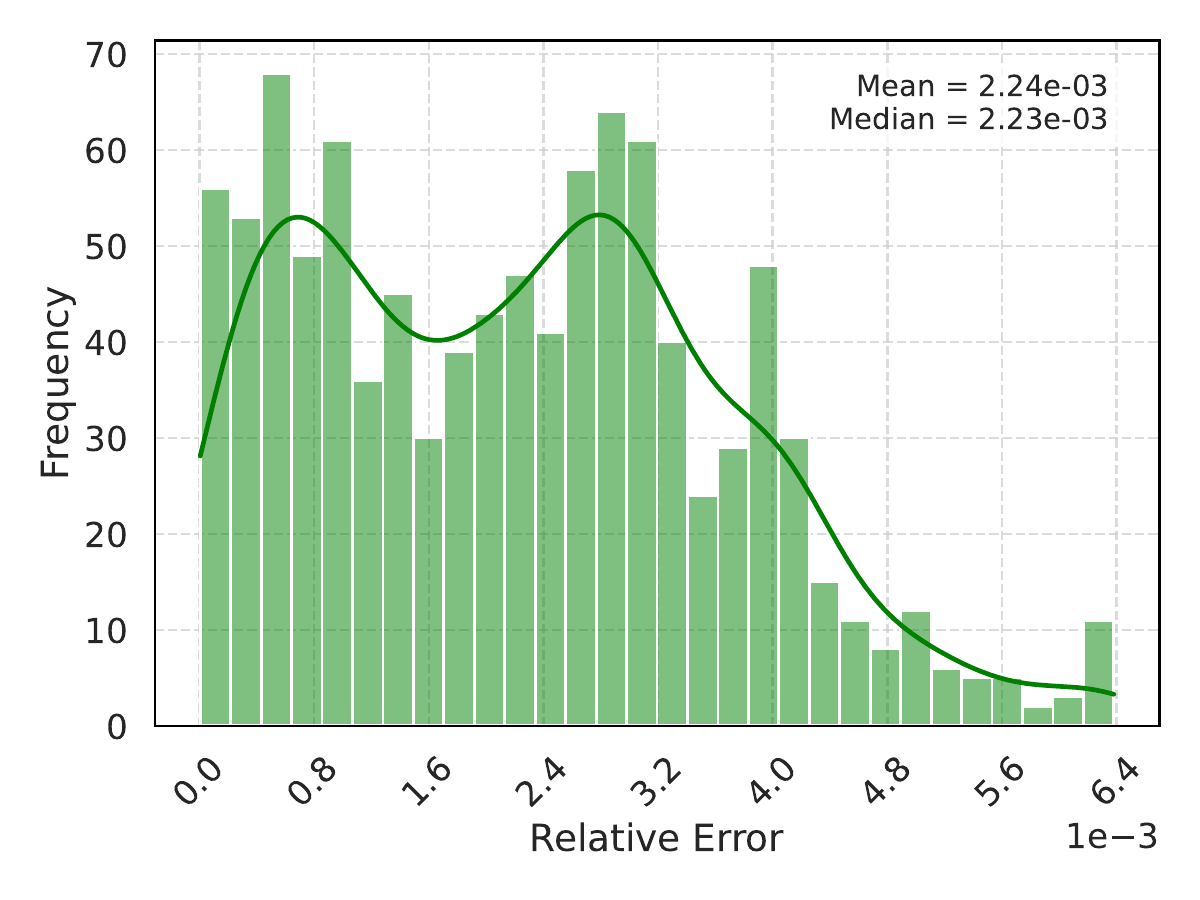}
    }\hfill
    \subfigure[Rel Error vs Freq. for 5th Order Taylor Tan]{
        \includegraphics[width=0.48\textwidth]{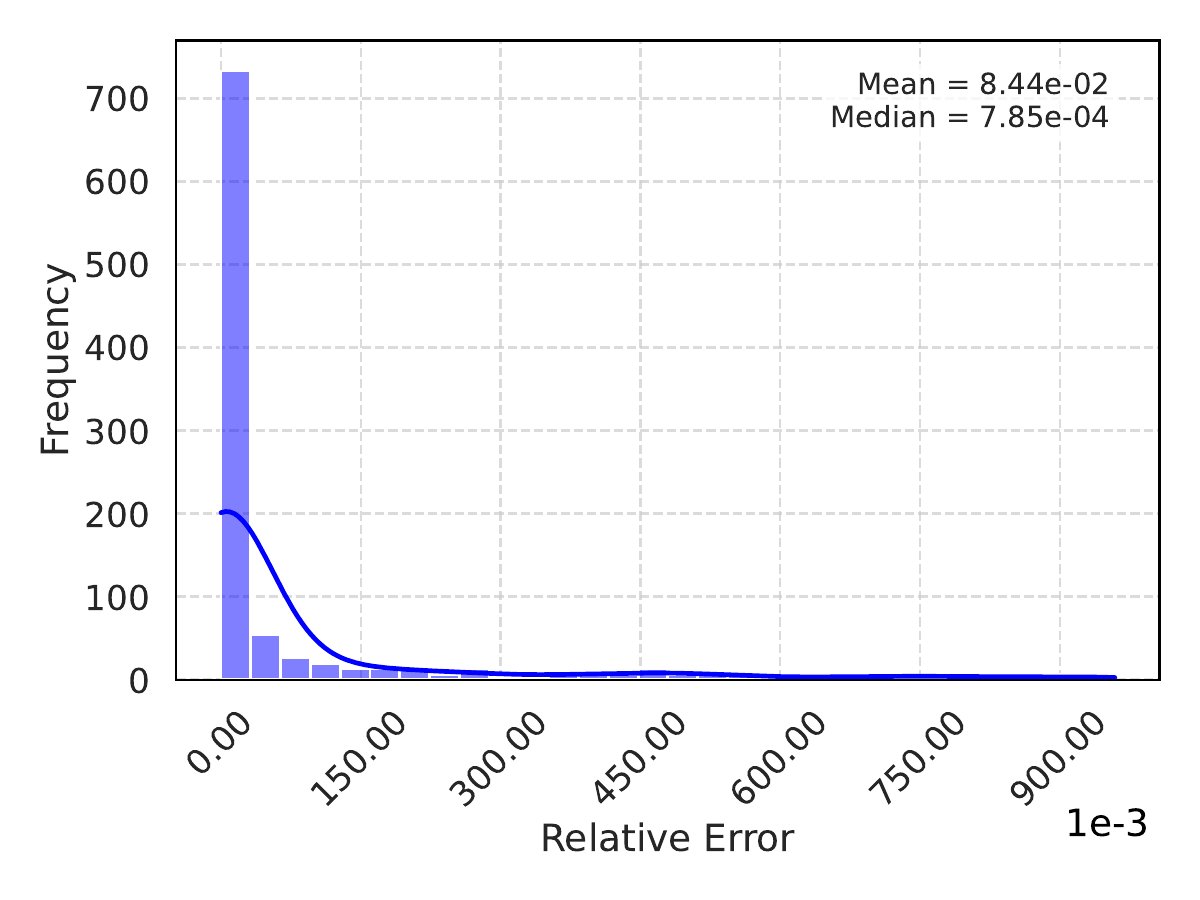}
    }
    \caption{Angle-wise Distribution of Relative error for tan}
    \label{rel_error_tan}
\end{figure}

\subsection{Device Usage Estimates}
As mentioned in the Introduction, SIMD optimizations are not directly available in Vitis HLS. Resulting in the inability of generating resource estimates for proposed as well as taylor in their most optimized form. However, by employing techniques like loop unrolling and pipelining, a good estimate of resource usage can be made.\\
As discussed in the Introduction, \textbf{Artix-7 XC7A35T-CPG236-1} was chosen for benchmarking to effectively evaluate the memory efficiency of the proposed formulas, particularly in resource-limited environments. Detailed hardware specifications can be found in the following section.
\subsubsection{Sin}

\begin{table}[H]
  \centering
  \caption{Resource Estimates for \texttt{Sin(double)}}
  \small
  \begin{tabular}{lccccc}
    \toprule
    Method                & BRAM       & DSP         & Flip-Flops           & LUT          \\
    \midrule
    Proposed              & 2 (2\%)    & 39 (43\%)   & 4461 (10\%)  & 5734 (27\%)     \\
    5-order Taylor Sin    & 2 (2\%)    & 14 (15\%)   & 2210 (5\%)   & 3349 (16\%)     \\
    Inbuilt Sin           & 8 (8\%)    & 85 (94\%)   & 4940 (11\%)  & 6641 (31\%)    \\
    \bottomrule
  \end{tabular}
  \label{tab:sin_performance}
\end{table}

\subsubsection{Cos}

\begin{table}[H]
  \centering
  \caption{Resource Estimates for \texttt{cos(double)}}
  \small
  \begin{tabular}{lccccc}
    \toprule
    Method                & BRAM       & DSP         & FF            & LUT          \\
    \midrule
    Proposed              & 2 (2\%)    & 39 (43\%)   & 4396 (10\%)   & 5633 (27\%)    \\
    5-order Taylor        & 2 (2\%)    & 14 (15\%)   & 2200 (5\%)    & 3208 (15\%)     \\
    Inbuilt               & 8 (8\%)    & 85 (94\%)   & 4940 (11\%)   & 6643 (31\%)    \\
    \bottomrule
  \end{tabular}
  \label{tab:cos_performance}
\end{table}

\subsubsection{Tan}
\begin{table}[H]
  \centering
  \caption{Resource Estimates for \texttt{tan(double)}}
  \small
  \begin{tabular}{lccccc}
    \toprule
    Method               & BRAM       & DSP           & FF           & LUT         \\
    \midrule
    Proposed             & 2 (2\%)    & 28 (31\%)     & 3944 (9\%)   & 5473 (26\%)   \\
    5-order Taylor tan   & 2 (2\%)    & 28 (31\%)     & 3283 (7\%)   & 4320 (20\%)   \\
    Inbuilt tan          & 8 (8\%)    & 111 (123\%)   & 7701 (18\%)  & 9181 (44\%)   \\
    \bottomrule
  \end{tabular}
  \label{tab:tan_performance}
\end{table}


It can be observed from Table's \ref{tab:sin_performance}, \ref{tab:cos_performance}, \ref{tab:tan_performance} that the proposed functions and  5th-order taylor functions demonstrate superior memory efficiency as compared to the inbuilt functions. Specifically :
\begin{enumerate}[label = \roman*]
    \item The proposed functions consume 4 times less BRAM across all cases. High BRAM usage can saturate memory resources especially on low-end FPGA's which limits their ability to contain multiple modules.
    \item The proposed functions use almost half the DSP's as compared to inbuilt functions and almost 4 times less DSP's in the case of tan. Marking a significant reduction in DSP usage which will allow it to run on resource constrained FPGA's.
    \item The proposed functions have lower Flip Flops and LUT usage for all cases. Notably for tan the proposed function uses just half as many FF's and LUT's as compared to the inbuilt function. This reduction in usage of FF's and LUT's results in low-power consumption which can be extremely useful for applications like IOT sensors and medical devices.  
\end{enumerate}

Clearly, taylor functions also exhibit excellent memory efficiency  performing equally well or better than the proposed functions in all three cases. Thus, both the proposed as well as the taylor functions are suitable candidates to be used in resource-constrained devices like low-end FPGA's and embedded systems. 
However, in the case of tan, taylor might not be a good choice because of its instability around odd multiples of $\frac{\pi}{2}$ which can cause problems in applications where high accuracy is essential, particularly on edge cases. 
In contrast, the proposed tan effectively addresses this issue as it maintains a high accuracy across all input ranges, including near vicinity values of odd multiples of $\frac{\pi}{2}$.

\section*{Hardware Specifications}
\label{hardware}

\begin{table}[H]
  \centering
  \caption{System and Device Specifications}
  \small
  \begin{tabular}{ll}
    \toprule
    \multicolumn{2}{l}{\textbf{Processor Specifications}} \\
    \midrule
    \textbf{Name} & AMD Ryzen 9 5900HX \\
    \textbf{Max Clock Speed (MHz)} & 3301 \\
    \textbf{Cores} & 8 \\
    \textbf{Logical Processors} & 16 \\
    \midrule
    \multicolumn{2}{l}{\textbf{Memory Specifications}} \\
    \midrule
    \textbf{Capacity (Bytes)} & 17179869184 + 17179869184 \\
    \textbf{Speed (MHz)} & 3200 \\
    \textbf{Manufacturer} & Samsung \\
    \midrule
    \multicolumn{2}{l}{\textbf{Video Controller Specifications}} \\
    \midrule
    \textbf{Name} & NVIDIA GeForce RTX 3080 Laptop GPU \\
    \textbf{Adapter RAM (Bytes)} & 4293918720 \\
    \textbf{Name} & AMD Radeon(TM) Graphics \\
    \textbf{Adapter RAM (Bytes)} & 536870912 \\
    \midrule
    \multicolumn{2}{l}{\textbf{Operating System Specifications}} \\
    \midrule
    \textbf{Windows Product Name} & Windows 10 Home Single Language \\
    \textbf{Windows Version} & 2009 \\
    \textbf{OS Architecture} & 64-bit \\
    \midrule
    \multicolumn{2}{l}{\textbf{FPGA Device Specifications}} \\
    \midrule
    \textbf{Part} & xc7a35tcpg236-1 \\
    \textbf{Family} & Artix-7 \\
    \textbf{Package} & cpg236 \\
    \textbf{Speed} & -1 \\
    \textbf{LUT} & 20800 \\
    \textbf{FF} & 41600 \\
    \textbf{DSP} & 90 \\
    \textbf{BRAM} & 50 \\
    \midrule
    \multicolumn{2}{l}{\textbf{Vitis and Project Configuration}} \\
    \midrule
    \textbf{Version} & 2024.2 (Build 5238294 on Nov 8 2024) \\
    \textbf{Project Name} & \texttt{trig\_approx} \\
    \textbf{Solution Type} & HLS (Vivado IP Flow Target) \\
    \textbf{Product Family} & artix7 \\
    \textbf{Target Device} & xc7a35t-cpg236-1 \\
    \bottomrule
  \end{tabular}
  \label{tab:hardware-specs}
\end{table}

\section{Conclusion}
In this paper comprehensive comparative benchmarks were run between proposed implementations and inbuilt library functions. The proposed implementations come out to be \(5 \times\) faster than its inbuilt functions, all while using no memory for precomputations. Notably, the proposed tangent function has been found to be extremely stable, even close to the singularity points of the tangent function, something that most approximation methods struggle with.
Further hardware usage benchmarks were conducted, which demonstrated a significant reduction in hardware resources such as DSPs, BRAMs, FFs, and LUTs compared to the inbuilt functions. This reduction highlights the suitability of the proposed implementations for resource-constrained devices, such as low-end FPGAs and MCU's.


\bibliographystyle{IEEEtran}  
\bibliography{references}  

\begin{thebibliography}{10}
\providecommand{\url}[1]{#1}
\csname url@samestyle\endcsname
\providecommand{\newblock}{\relax}
\providecommand{\bibinfo}[2]{#2}
\providecommand{\BIBentrySTDinterwordspacing}{\spaceskip=0pt\relax}
\providecommand{\BIBentryALTinterwordstretchfactor}{4}
\providecommand{\BIBentryALTinterwordspacing}{\spaceskip=\fontdimen2\font plus
\BIBentryALTinterwordstretchfactor\fontdimen3\font minus \fontdimen4\font\relax}
\providecommand{\BIBforeignlanguage}[2]{{%
\expandafter\ifx\csname l@#1\endcsname\relax
\typeout{** WARNING: IEEEtran.bst: No hyphenation pattern has been}%
\typeout{** loaded for the language `#1'. Using the pattern for}%
\typeout{** the default language instead.}%
\else
\language=\csname l@#1\endcsname
\fi
#2}}
\providecommand{\BIBdecl}{\relax}
\BIBdecl

\bibitem{r1}
N.~Ramya and T.~Vijayaram, ``Applications of trigonometry in engineering and technology,'' \emph{International Journal of Structural Mechanics and Finite Elements}, vol.~10, no.~1, pp. 11--15, 2024.

\bibitem{taylorser}
\BIBentryALTinterwordspacing
H.~Haber, ``Taylor series expansions,'' 2011, accessed: 2025-02-13. [Online]. Available: \url{https://scipp.ucsc.edu/~haber/ph116A/taylor11.pdf}
\BIBentrySTDinterwordspacing

\bibitem{cordic_1}
R.~Andraka, ``A survey of cordic algorithms for fpga based computers,'' \emph{ACM/SIGDA International Symposium on Field Programmable Gate Arrays - FPGA}, 12 2001.

\bibitem{cordic_2}
L.~Chen, J.~Han, W.~Liu, and F.~Lombardi, ``Algorithm and design of a fully parallel approximate coordinate rotation digital computer (cordic),'' \emph{IEEE Transactions on Multi-Scale Computing Systems}, vol.~3, no.~3, pp. 139--151, 2017.

\bibitem{lut_1}
\BIBentryALTinterwordspacing
{MATLAB}, ``Optimize lookup tables for memory-efficiency programmatically,'' 2021, accessed: 2025-02-13. [Online]. Available: \url{https://de.mathworks.com/help/fixedpoint/ug/optimize-lookup-tables-for-memory-efficiency-programmatically.html}
\BIBentrySTDinterwordspacing

\bibitem{lut_2}
Y.~Tian, T.~Wang, Q.~Zhang, and Q.~Xu, ``Approxlut: a novel approximate lookup table-based accelerator,'' in \emph{Proceedings of the 36th International Conference on Computer-Aided Design}, ser. ICCAD '17.\hskip 1em plus 0.5em minus 0.4em\relax IEEE Press, 2017, p. 438–443.

\bibitem{poly1}
B.~Adcock, S.~Brugiapaglia, and C.~Webster, ``Compressed sensing approaches for polynomial approximation of high-dimensional functions,'' \emph{Applied and Numerical Harmonic Analysis}, 03 2017.

\bibitem{poly2}
\BIBentryALTinterwordspacing
J.~T. Butler, C.~Frenzen, N.~Macaria, and T.~Sasao, ``A fast segmentation algorithm for piecewise polynomial numeric function generators,'' \emph{Journal of Computational and Applied Mathematics}, vol. 235, no.~14, pp. 4076--4082, 2011. [Online]. Available: \url{https://www.sciencedirect.com/science/article/pii/S037704271100121X}
\BIBentrySTDinterwordspacing

\bibitem{poly3}
D.~De~Caro, N.~Petra, and A.~G.~M. Strollo, ``Efficient logarithmic converters for digital signal processing applications,'' \emph{IEEE Transactions on Circuits and Systems II: Express Briefs}, vol.~58, no.~10, pp. 667--671, 2011.

\bibitem{poly4}
H.~Dong, M.~Wang, Y.~Luo, M.~Zheng, M.~An, Y.~Ha, and H.~Pan, ``Plac: Piecewise linear approximation computation for all nonlinear unary functions,'' \emph{IEEE Transactions on Very Large Scale Integration (VLSI) Systems}, vol.~PP, pp. 1--14, 07 2020.

\bibitem{gitcode}
\BIBentryALTinterwordspacing
N.~Goyal, ``trigno\_simd\_cpp: C++ implementations of trigonometric functions with simd optimization,'' 2025, gitHub repository. [Online]. Available: \url{https://github.com/Nikhil0250/trigno_simd_cpp}
\BIBentrySTDinterwordspacing

\bibitem{fisr}
\BIBentryALTinterwordspacing
C.~J. Walczyk, L.~V. Moroz, and J.~L. Cieśliński, ``Improving the accuracy of the fast inverse square root by modifying newton–raphson corrections,'' \emph{Entropy}, vol.~23, no.~1, 2021. [Online]. Available: \url{https://www.mdpi.com/1099-4300/23/1/86}
\BIBentrySTDinterwordspacing

\bibitem{tayloropt1}
Z.~Duoli, J.~Yu, and Y.~Song, ``Implementation of high accuracy trigonometric function on fpga by taylor expansion,'' 01 2016.

\bibitem{tayloropt2}
\BIBentryALTinterwordspacing
T.~Kusaka and T.~Tanaka, ``Fast and accurate approximation methods for trigonometric and arctangent calculations for low-performance computers,'' \emph{Electronics}, vol.~11, no.~15, 2022. [Online]. Available: \url{https://www.mdpi.com/2079-9292/11/15/2285}
\BIBentrySTDinterwordspacing

\end{thebibliography}

\appendix

\section{Mathematical Proofs}
\label{appendix:proofs}
As discussed in Section \ref{sec:proposed}, linear interpolation for sin and cos will be used for the derivation of tan formulas. Subsequently these tan functions will be used to derive the necessary functions for sin and cos. \textbf{The expressions obtained here are derived using degrees as the unit. These formulas can be converted to the ones used in the paper by applying a standard conversion to radians}
\subsection{Linear Interpolation of sine}
The methodology is to first interpolate $sinx^\circ$ as a straight line for $15^\circ \leq x \leq 75^\circ$ at an interval of 15 degrees. 
\\
The equations for each interval will be written utilizing the two-point form which is : \\
$$
y - y_1=(x-x_1)\frac{y_2-y_1}{x_2 - x_1} \quad \text{where both $(x_1,y_1)$ and $(x_2, y_2)$ lie on the line}
$$
\\
\textbf{For : $15^\circ \leq x \leq 30^\circ$ }
\begin{align}
sin~x &= sin~15^\circ +\frac{(x-15^\circ)(sin~30^\circ -sin~15^\circ)}{30-15} \nonumber
\\
\text{Upon further solving,} \nonumber
\\
sin~x &\approx \frac{x(sin~30^\circ - sin~15^\circ) + 30sin~15^\circ -15 sin~30^\circ}{15}  \label{1}
\end{align}
\textbf{For : $30^\circ \leq x \leq 45^\circ$ }\\
\begin{align}
sin~x &= sin~30^\circ +\frac{(x-30^\circ)(sin~45^\circ -sin~30^\circ)}{45-30} \nonumber
\\
\text{Upon further solving,} \nonumber
\\
sin~x &\approx \frac{x(sin~45^\circ - sin~30^\circ) + 45sin~30^\circ -30 sin~45^\circ}{15} \label{2}
\end{align}
\textbf{For : $45^\circ \leq x \leq 60^\circ$ }\\
\begin{align}
sin~x &= sin~45^\circ +\frac{(x-45^\circ)(sin~60^\circ -sin~45^\circ)}{60-45} \nonumber
\\
\text{Upon further solving,} \nonumber
\\
sin~x &\approx \frac{x(sin~60^\circ - sin~45^\circ) + 60sin~45^\circ -45 sin~60^\circ}{15} \label{3}
\end{align}
\textbf{For : $60^\circ \leq x \leq 75^\circ$ }\\
\begin{align}
sin~x &= sin~60^\circ +\frac{(x-60^\circ)(sin~75^\circ -sin~60^\circ)}{75-60} \nonumber
\\
\text{Upon further solving,} \nonumber
\\
sin~x &\approx \frac{x(sin~75^\circ - sin~60^\circ) + 75sin~60^\circ -60 sin~75^\circ}{15} \label{4}
\end{align}
\subsection{\texorpdfstring{$\tan x$ for $15^\circ$ to $75^\circ$}{tan x for 15 degrees to 75 degrees}}

Now, interpolation formulas for sine obtained in the previous section will be used to arrive at the results for $tanx$ ranging from $15^\circ$ to $75^\circ$. \\
The methodology will be to write $tan~x=\frac{sin~x}{sin~(90^\circ-x)}$ and then use the interpolated equations previously obtained for the corresponding ranges of $x$ and $(90-x)$.\\
\textbf{For : $15^\circ \leq x \leq 30^\circ$ }\\
\vspace{1.5pt}
Here, $60^\circ \leq (90^\circ-x) \leq 75^\circ$. From \eqref{1} and \eqref{4}\\
\begin{align*}
sin~x &= \frac{x(sin~30^\circ - sin~15^\circ) + 30sin~15^\circ -15 sin~30^\circ}{15}\\
sin~(90^\circ-x) &= \frac{(90^\circ-x)(sin~75^\circ - sin~60^\circ) + 75sin~60^\circ -60 sin~75^\circ}{15}\\
\frac{sin~x}{sin~(90^\circ-x)}&=tan~x \approx \frac{x(sin~30^\circ - sin~15^\circ) + 30sin~15^\circ -15 sin~30^\circ}{(90-x)(sin~75^\circ - sin~60^\circ) + 75sin~60^\circ -60 sin~75^\circ}
\end{align*}
\\
\vspace{1.5pt}
Simplifying,
\\
\vspace{1.5pt}
\begin{align}
tan~x \approx  \frac{0.2411x+0.2645}{-0.099x+15.9673} \approx \frac{10x+10}{-4x+657} \label{5}
\end{align}
Converting to radians, the desired formula is obtained, 
\\
\vspace{1.5pt}
\begin{align}
tan~x \approx \frac{572.95 x + 10.0}{-229.18 x + 657.0}
\end{align}
\\
\textbf{For : $30^\circ \leq x \leq 45^\circ$ }\\
\vspace{1.5pt}
Here, $45^\circ \leq (90^\circ-x) \leq 60^\circ$. From \eqref{2} and \eqref{3}\\
\begin{align*}
sin~x &= \frac{x(sin~45^\circ - sin~30^\circ) + 45sin~30^\circ -30 sin~45^\circ}{15}\\
sin~(90^\circ-x) &= \frac{(90^\circ-x)(sin~60^\circ - sin~45^\circ) + 60sin~45^\circ -45 sin~60^\circ}{15}\\
\frac{sin~x}{sin(90^\circ-x)}&=tan~x \approx \frac{x(sin~45^\circ - sin~30^\circ) + 45sin~30^\circ -30 sin~45^\circ}{(90-x)(sin~60^\circ - sin~45^\circ) + 60sin~45^\circ -45 sin~60^\circ}
\end{align*}
\vspace{1.5pt}
Simplifying,
\\
\vspace{1.5pt}
\begin{align}
tan~x \approx \frac{0.2071x+1.2867}{-0.1589x+17.7579} \approx \frac{6x+46}{-5x+541} \label{6}
\end{align}
Converting to radians, the desired formula is obtained:
\\
\vspace{1.5pt}
\begin{align}
tan~x \approx \frac{343.77 x + 46.0}{-286.47 x + 541.0} 
\end{align}
\\
\textbf{For $45^\circ \leq x \leq 60^\circ$ }\\
\vspace{1.5pt}
Here, $30^\circ \leq (90-x) \leq 45^\circ$. From \eqref{3} and \eqref{2}\\
\begin{align*}
sin~x &= \frac{x(sin~60^\circ - sin~45^\circ) + 60sin~45^\circ -45 sin~60^\circ}{15}\\
sin~(90^\circ-x) &= \frac{(90^\circ-x)(sin~45^\circ - sin~30^\circ) + 45sin~30^\circ -30 sin~45^\circ}{15}\\
\frac{sin~x}{sin~(90^\circ-x)} &= tan~x \approx \frac{x(sin~60^\circ - sin~45^\circ) + 60sin~45^\circ -45 sin~60^\circ}{(90-x)(sin~45^\circ - sin~30^\circ) + 45sin~30^\circ -30 sin~45^\circ} 
\end{align*}
\vspace{1.5pt}
Simplifying,
\\
\vspace{1.5pt}
\begin{align}
tan~x \approx \frac{0.1589x+3.4552}{-0.2071x+19.9264}\approx \frac{10x+217}{-13x+1252} \label{7}
\end{align}
Converting to radians, the desired formula is obtained:
\\
\vspace{1.5pt}
\begin{align}
tan~x \approx \frac{572.95 x + 217.0}{-744.84 x + 1252.0}
\end{align}
\\
\textbf{For $60^\circ \leq x \leq 75^\circ$ }\\
\vspace{1.5pt}
Here, $15^\circ \leq (90-x) \leq 30^\circ$. From \eqref{4} and \eqref{1}\\
\begin{align*}
sin~x &= \frac{x(sin~75^\circ - sin~60^\circ) + 75sin~60^\circ -60 sin~75^\circ}{15}\\
sin~(90^\circ-x) &= \frac{(90^\circ-x)(sin~30^\circ - sin~15^\circ) + 30sin~15^\circ -15 sin~30^\circ}{15}\\
\frac{sin~x}{sin~(90^\circ-x)} &= tan~x \approx \frac{x(sin~75^\circ - sin~60^\circ) + 75sin~60^\circ -60 sin~75^\circ}{(90-x)(sin~30^\circ - sin~15^\circ) + 30sin~15^\circ -15 sin~30^\circ}
\end{align*}
\vspace{1.5pt}
Simplifying,
\\
\vspace{1.5pt}
\begin{align}
tan~x \approx \frac{0.0999x+6.9963}{-0.2411x+21.9708} \approx \frac{297+4x}{-10x+910} \label{8}
\end{align}
Converting to radians, the desired formula is obtained:
\\
\vspace{1.5pt}
\begin{align}
tan~x \approx \frac{229.18 x + 297.0}{-572.95 x + 910.0}
\end{align}
\subsection{\texorpdfstring{$\tan x$ for $0^\circ$ to $15^\circ$ and $75^\circ$ to $90^\circ$}{tan x for 0 degrees to 15 degrees and 75 degrees to 90 degrees}}

The formulas obtained in the previous section will be used to obtain formulas for tan in the range $0^\circ \leq x \leq 15^\circ$ and $75^\circ \leq x < 90^\circ$.\\
The following identities are recalled : \\
$$tan~(a-b)=\frac{tan~a-tan~b}{1+tan~atan~b}$$
\\
$$tan~(a+b)=\frac{tan~a+tan~b}{1-tan~atan~b}$$
\\
To calculate $tan~x^\circ$ for $0^\circ \leq x \leq 15^\circ$. The expression \( \tan(45^\circ - x') \) can be used, where \(30^\circ \leq x' \leq 45^\circ\): \\
$$tan~(45^\circ-x')=\frac{1-tan~x'}{1+tan~x'}$$
\\
Substituting $tan~x'$ from \eqref{6}
$$tan~(45^\circ-x')\approx \frac{1-\frac{6x'+46}{541-5x'}}{1+\frac{6x'+46}{541-5x'}}$$
\\
\vspace{1.5pt}
Simplifying,
\\
\vspace{1.5pt}
$$tan~(45^\circ-x')\approx \frac{495-11x'}{587+x'}$$
\\
Now putting $(45-x')$ as $x$ because $0^\circ \leq (45-x') \leq 15^\circ$:
\\
\begin{align}
tan~x \approx \frac{11x}{-x+632} \label{9}
\end{align}
\\
Converting to radians, the desired formula is obtained: 
\\
\vspace{1.5pt}
\begin{align}
tan~x \approx \frac{630.25 x}{-57.29 x + 632.0}
\end{align}
To calculate $tan~x$ for $75^\circ \leq x < 90^\circ$. The expression \( \tan(45^\circ + x') \) can be used, where \(30^\circ \leq x' < 45^\circ\): \\

$$tan~(45^\circ+x')=\frac{1+tan~x'}{1-tan~x'}$$
\\
Substituting $tan~x'$ from \eqref{6}
\\
$$tan~(45^\circ+x')\approx \frac{1+\frac{6x'+46}{541-5x'}}{1-\frac{6x'+46}{541-5x'}}$$
\\
\vspace{1.5pt}
Simplifying,
\\
\vspace{1.5pt}
$$tan~(45^\circ+x')\approx\frac{587+x'}{495-11x'}$$
\\
Now putting $(45+x')$ as $x$ because $75^\circ \leq (45+x') < 90^\circ$:
\\
\begin{align}
tan~x \approx \frac{542+x}{-11x+990} \label{10}
\end{align}
Converting to radians, the desired formula is obtained: 
\\
\vspace{1.5pt}
\begin{align}
tan~x \approx \frac{57.29 x + 542.0}{-630.25 x + 990.0}
\end{align}

\subsection{Formulas for sin and cos}
The desired expressions for sin and cos can be obtained from tan by realizing some standard trigonometric identities.\\
Since the expressions for tan are of the form: 
\vspace{1.5pt}
\begin{align}
tan~x \approx \frac{ax + b}{cx + d}
\end{align}
General expression for sin and cos can be written as :\\
\vspace{1.5pt}
\begin{align}
sin~x \approx \frac{ax + b}{\sqrt{(a^2 + c^2)x^2 + 2(ab + cd)x + (b^2 + d^2)}}
\end{align}

\vspace{1.5pt}
\begin{align}
cos~x \approx \frac{cx + d}{\sqrt{(a^2 + c^2)x^2 + 2(ab + cd)x + (b^2 + d^2)}}
\end{align}
By substituting coefficients a, b, c and d accordingly, the expressions used in the paper can be obtained.

\label{'ubl'}  
\end{document}